\documentclass[journal,letterpaper,12pt,oneside,onecolumn,draftclsnofoot]{IEEEtran}

\usepackage[cmex10]{amsmath}
\usepackage{amsfonts}
\usepackage{mathrsfs}
\usepackage{txfonts}
\usepackage{amssymb}   % From the American Mathematical Society
\usepackage{mathrsfs}
\usepackage{algorithm}
\usepackage{algorithmic}
\usepackage{color}
\usepackage{cite}      % Written by Donald Arseneau
\usepackage{graphicx}
\usepackage{psfrag}    % Written by Craig Barratt, Michael C. Grant,
\usepackage{subfigure} % Written by Steven Douglas Cochran
\usepackage{url}       % Written by Donald Arseneau
\usepackage{stfloats}  % Written by Sigitas Tolusis
\usepackage{enumerate}
\usepackage{multirow}
\usepackage{booktabs}
\usepackage{setspace}
\usepackage{longtable}
\usepackage{threeparttable}
\usepackage{makecell}

\newcommand{\qed}{\nobreak \ifvmode \relax \else
	\ifdim\lastskip<1.5em \hskip-\lastskip
	\hskip1.5em plus0em minus0.5em \fi \nobreak
	\vrule height0.75em width0.5em depth0.25em\fi}

\begin{document}

\title{Software Defined Space-Air-Ground Integrated Vehicular Networks: Challenges and Solutions}
\author{\small Ning Zhang,
               Shan Zhang,
               Peng Yang,
               Omar Alhussein,
               Weihua Zhuang,
               Xuemin~(Sherman)~Shen

\thanks{Ning Zhang, Shan Zhang, Omar Alhussein, Weihua Zhuang, and  Xuemin (Sherman) Shen are with University of Waterloo, Canada; and Peng Yang is with Huazhong University of Science and Technology, China.}}
%\thanks{Dajiang Chen is with School of information and software engineering, University of Electronic Science and Technology of
%China, Chengdu, China (e-mail: djchen@uestc.edu.cn).}
%\thanks{Peng Yang is with Huazhong University of Science and Technology, China}
%\thanks{Ben Liang is with University of Toronto, Canada.}
%}

%\thanks{Ning Zhang, Shan Zhang, and Xuemin (Sherman) Shen are with the Department of Electrical and Computer Engineering, University of Waterloo, Canada (e-mail:\{n35zhang, sshen\}@uwaterloo.ca; zhangshan2007@gmail.com).}
%\thanks{Dajiang Chen is with School of information and software engineering, University of Electronic Science and Technology of
%China, Chengdu, China (e-mail: djchen@uestc.edu.cn).}
%\thanks{Peng Yang is with Huazhong University of Science and Technology, China}
%\thanks{Ben Liang is with University of Toronto, Canada.}
%}
\maketitle
%\begin{abstract}
%%\boldmath
%\end{abstract}
%
%\begin{IEEEkeywords}
%\end{IEEEkeywords}

\IEEEpeerreviewmaketitle

\begin{abstract}
This article proposes a software defined space-air-ground integrated network architecture for supporting diverse vehicular services in a seamless, efficient, and cost-effective manner. Firstly, the motivations and challenges for integration of space-air-ground networks are reviewed. Secondly, a software defined network architecture with a layered structure is presented. To protect the legacy services in satellite, aerial, and territorial segments, resources in each segment are sliced through network slicing to achieve service isolation. Then, available resources are put into a common and dynamic space-air-ground resource pool, which is managed by hierarchical controllers to accommodate vehicular services. Finally, a case study is carried out, followed by discussion on some open research topics.

\end{abstract}
\section{Introduction}

The connected vehicle paradigm is to empower vehicles to communicate with the surrounding environment such as neighboring cars, road-side infrastructure, and traffic control centers, playing a vital role in next generation intelligent transportation (ITS) system. With connected vehicles, a wide range of on-the-go services can be facilitated, including road safety (e.g., collision avoidance and intelligent traffic management), infotainment (e.g., social networking and online gaming), and location dependent services (e.g., point of interests and route optimization) \cite{lu2014connected,abboudinterworking,cheng2016opportunistic}. With its great potential, connected vehicle is regarded as the next frontier for automotive revolution and the number of connected vehicles is predicted to reach 250 million by 2020\footnote{Gartner,``Connected cars from a major element of internet of things," http://www.gartner.com/newsroom/id/2970017, 2015.}.

Extensive research and development efforts have been made from both industry and academia to get vehicles connected. The main enabling platforms include Dedicated Short-Range Communications (DSRC) based 802.11p networks and cellular networks \cite{schwarz2016society}. The former facilitates both vehicle-to-vehicle (V2V) and vehicle-to-roadside (V2R) communications, while the latter such as Long-Term Evolution (LTE) can provide reliable access to the Internet. However, for 802.11p networks, it is costly and will take a long time to deploy the large-scale network infrastructure such as roadside units (RSUs). Although the LTE network can exploit the existing infrastructure, it faces the issues of coverage and capacity. For instance, the LTE network has poor coverage in rural areas, while having potential network congestion in urban areas due to ever-increasing wireless devices and services \cite{amadeo2011satellite}. Furthermore, both 802.11p and cellular networks have great challenges to support high mobility, and highly mobile vehicles can suffer from frequent handovers to associate with new RSUs and base stations respectively, as the networks are becoming even denser. %As the mobile networks become more dense to enhance network capacity, the high mobile vehicles can suffer from frequent handover due to the shrunken cell size.

In addition to terrestrial networks, other complementary solutions, such as satellite and aerial networks, are under development. For example, Intelsat satellite antenna (i.e., mTenna) can be embedded into the roof of the cars to acquire satellite signals. Toyota Mirai Research Vehicle equipped with mTenna can provide on-the-move services, which is demonstrated to achieve a data rate of 50 megabits per second. In addition, Tesla plans to launch 700 low-cost commercial satellites to provide Internet access, while Google will launch a fleet of 180 satellites\footnote{http://www.dailymail.co.uk/sciencetech/article-2828539/Elon-Musk-s-mission-700-cheap-satellites-provide-internet-access-world.html, 2014.}. In parallel, high-altitude platforms (HAPs) can also provide Internet access to vehicles \cite{zhou2015multi}. As a matter of fact, Project Loon initiated by Google aims to leverage high-altitude balloons for broadband services in remote locations, and Facebook attempts to deploy solar-powered drones to provide Internet access to underdeveloped areas.

To accommodate the diverse vehicular services with different quality of service (QoS) requirements in various practical scenarios (e.g., rural and urban), it is imperative to exploit specific advantages of each networking paradigm. For instance, densely deployed terrestrial networks in urban areas can support high-data rate access, satellite communication systems can provide seamless connectivity to rural areas, while HAPs can enhance the capacity for areas with high service demands. In addition, multi-dimensional real-time context-aware information regarding the vehicular environments, such as in-vehicle, inter-vehicle, road conditions, and regional transport information, can be acquired to improve driving experience and facilitate intelligent traffic management. In this article, we focus on space-air-ground integrated vehicular networks, by means of software defined networking (SDN)\cite{kim2013improving}, to agilely and flexibly exploit heterogenous resources to support heterogenous vehicular services. The motivation and challenges for space-air-ground integrated networks are first discussed in Section~\ref{sec.Evolution}. Then, a software defined space-air-ground integrated vehicular network architecture is proposed in Section~\ref{sec.SDN}. In Section~\ref{sec.operation}, the working relation is presented, along with the hierarchical network operation and big data-assisted networking. In Section~\ref{sec.Research}, the research directions are identified, followed by the conclusion of this work in Section~\ref{sec.Conclusion}.

\section{Space-Air-Ground Integrated Vehicular Network: Motivations and Challenges}\label{sec.Evolution}

\subsection{Background}
%Entities, links, spectrum, main services, ....
A satellite network is composed of a number of satellites, ground stations (GSs), and network operations control centers (NOCCs), and usually provides services for navigation, earth observation, emergency rescue, and communication/relaying. Based on the altitude, satellites can be categorized into geostationary orbit (GSO), medium earth orbit (MEO), and low earth orbit (LEO) satellites. Satellite networks have the advantages of: i) large coverage (3 GSO satellites or a constellation of LEO satellites such as Iridium composed of 77 LEO satellites can provide the global coverage) ; ii) broadcasting/multicasting capability to support a large number of users simultaneously; and iii) reliable access for extreme scenarios such as in disaster relief. In addition, broadband satellite systems are expected to have capacity of 1000 Gbps by the year of 2020 \cite{gharanjik2015multiple}. However, the round-trip delay especially with GSO and MEO satellites is relatively large, which is not suited for realtime applications. %Spectrum in satellite networks includes s, ku, ka bands.
%\cite{Yu2016}

%\cite{hu2001satellite}

An aerial network formed by quasi-stationary HAPs (17-22 km above the earth) in the stratosphere can also be employed to provide broadband connectivity. HAPs mainly consist of unmanned airships (e.g., balloons) and aircrafts such as unmanned aerial vehicles (UAVs). The solar-powered airships can stay in the air for around 5 years. Compared with base stations in terrestrial communication networks, HAPs can have a large coverage to offer services on a regional basis, and their movement can be partially controlled on demand to provide supplemental capacity in hot-spot areas. Moreover, HAPs have the features of low cost, as well as incremental and easy deployment.

For terrestrial networks, a dedicated spectrum with bandwidth 75 MHz at 5.9 GHz has been allocated for DSRC, while cellular networks are now evolving to next generation (5G) wireless networks, to support diverse services including vehicular communications. As for standardization, the Third Generation Partnership Project (3GPP) aims to develop a set of LTE specifications for vehicular environments (LTE-V). The terrestrial networks can provide high data rates to users, but the network coverage in rural and remote areas is poor and the capacity in hot-spots needs to be enhanced. A comparison of the different networking paradigms is given in Table I.

%\begin{table*}[t]\small
%\caption{A comparison of different paradigms.}\label{tab:notation}
%%\centering
%\begin{tabular}{|c|c|c|c|c|c|c|}
%\hline
% Segment &Entities &Altitude &Mobility	&Roundtrip  &Advantages	&Limitation \\ \hline
%\multirow{3}*{Space} & GEO	&35,786km  &Static to earth  &250-280 ms  &\multirow{3}{3.5cm}{Broadcast/multicast,\\ large coverage, rapid commercialization.} &\multirow{3}{4cm}{Long propagation delay; limited capacity, least flexibility, costly.} \\ \cline{2-5}
%                         & MEO	&3000km 	&Medium fast       &110-130 ms  &                     &           \\ \cline{2-5}
%                         & LEO	&200-3000km	&Fast          &20-25 ms    &                     &          \\ \hline
% \multirow{2}*{Air} & HAP	&17-22 km &Quasi-stationary &\multirow{2}*{Medium} &\multirow{2}{3.5cm}{Large coverage, low cost, flexible movement.
%} &\multirow{2}{4cm}{Less capacity, link instability.} \\ \cline{2-4}
%                    & UAV	&Up to 30km  &Varying speeds & & & \\ \hline
% \multirow{2}*{Ground} & DSRC	&N.A. &Static &\multirow{2}*{Lowest} &\multirow{2}{3.5cm}{Rich resources, high throughput.} &\multirow{2}{4cm}{Coverage, mobility support, vulnerable to disaster.\\} \\ \cline{2-4}
%                    & Celluar	&N.A. &Static & & & \\ \hline
%\end{tabular}
%\end{table*}
\subsection{Motivations for Integration}

\begin{table*}[t]\small
\caption{A comparison of different paradigms.}\label{tab:notation}
%\centering
\begin{tabular}{|c|c|c|c|c|c|c|}
\hline
 Segment &Entities &Altitude &Mobility	&Roundtrip  &Advantages	&Limitation \\ \hline
\multirow{3}*{Space} & GEO	&35,786km  &Static to earth  &250-280 ms  &\multirow{3}{3.5cm}{Broadcast/multicast,\\ large coverage, rapid commercialization.} &\multirow{3}{4cm}{Long propagation delay; limited capacity, least flexibility, costly.} \\ \cline{2-5}
                         & MEO	&3000km 	&Medium fast       &110-130 ms  &                     &           \\ \cline{2-5}
                         & LEO	&200-3000km	&Fast          &20-25 ms    &                     &          \\ \hline
 \multirow{2}*{Air} & HAP	&17-22 km &Quasi-stationary &\multirow{2}*{Medium} &\multirow{2}{3.5cm}{Large coverage, low cost, flexible movement.
} &\multirow{2}{4cm}{Less capacity, link instability.} \\ \cline{2-4}
                    & UAV	&Up to 30km  &Varying speeds & & & \\ \hline
 \multirow{2}*{Ground} & DSRC	&N.A. &Static &\multirow{2}*{Lowest} &\multirow{2}{3.5cm}{Rich resources, high throughput.} &\multirow{2}{4cm}{Coverage, mobility support, vulnerable to disaster.\\} \\ \cline{2-4}
                    & Celluar	&N.A. &Static & & & \\ \hline
\end{tabular}
\end{table*}

From a service perspective, different QoS requirements imposed by diverse services should be satisfied in various vehicular networking scenarios in a cost-effective and flexible manner. However, the standalone territorial network has many challenges to meet such needs:
\begin{itemize}
  \item The coverage of territorial networks in rural areas (e.g., mountain areas) and highways is poor. It is very costly to deploy more network infrastructure in those sparsely populated areas. Instead, existing satellite communication systems or HAPs can efficiently provide vehicular connectivity due to their large coverage. For instance, a satellite cell diameter can be several hundred kilometers, which is equivalent to several thousands of terrestrial base stations each having a coverage area of several kilometers in diameter \cite{kawamoto2014prospects}; %\cite{pace2004integrated}
  \item The moving vehicles suffer from frequent handover in territorial networks, degrading the on-the-move service performance. The coverage diameter of an LTE macrocell base station (BS) is around 1km, while the coverage diameter of a small cell BS (SBS) is less than 300m \cite{lopez2011enhanced}. Consequently, the moving vehicles on highways will experience highly frequent handovers from BS to BS;
  \item The diverse vehicular services cannot be served efficiently by a single technology. The LTE network can provide a high-data rate for individual vehicular users, but it is not designed to efficiently broadcast similar contents to a large number of users. In such a case, satellite/HAP communication systems are more efficient, thanks to their efficient broadcasting and multicasting capacity;
  \item With high spatial-temporal dynamics in traffic loads due to vehicle mobility, network congestion can occur even in urban/suburban areas. However, it is not cost-effective to add more territorial network resources to meet the peak traffic demands. Exploiting available resources from other systems such as HAPs can be more effective to accommodate the bursty traffic demands.
\end{itemize}

 Through interworking, the advantages from different segments can be exploited to support multifarious vehicular services and scenarios in an efficient and cost-effective manner. For instance, territorial networks can serve individual vehicular users through high-rate unicast in urban/suburban areas, while satellite networks help achieve ubiquitous coverage in rural and remote areas. Low-cost HAPs can be utilized to boost capacity at areas with poor or congested terrestrial infrastructure deployment; they can be repositioned to provide emergency communications in disaster scenarios. Moreover, both satellites and HAPs can provide road information and geolocation information to assist terrestrial networks, facilitate information dissemination as relays, and relieve the demands on terrestrial networks through data offloading.

\begin{figure*}[t!]
    \centering
    \includegraphics[width=12cm]{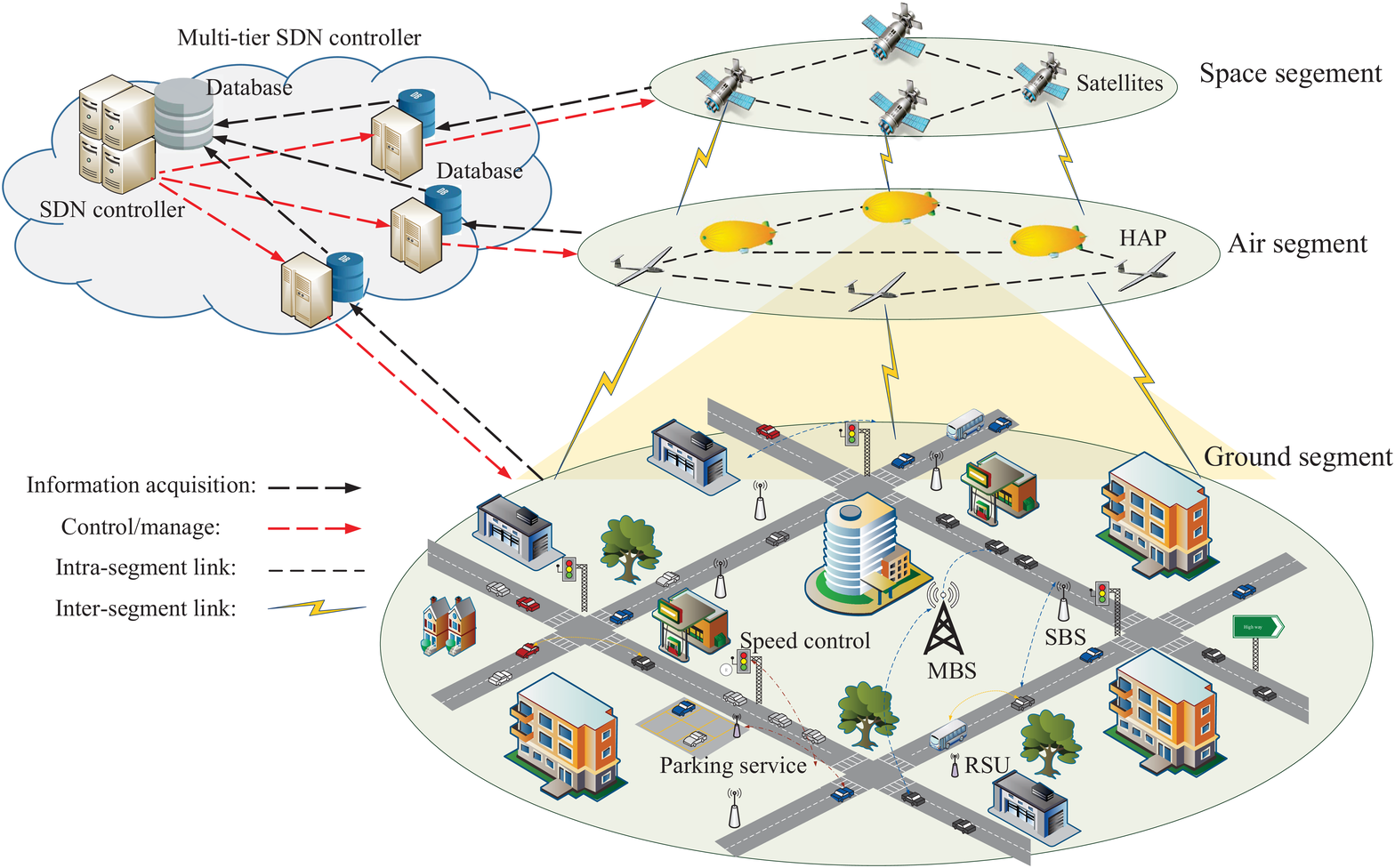}
    \caption{Software Defined Space-Air-Ground Intergraded Vehicular (SSAGV) Networks.}
    \label{fig:ssag}
\end{figure*}

\subsection{Challenges}
To integrate the space-air-ground networks for vehicular services, there exist many challenges:
\begin{itemize}
\item Inter-operation:
  Different networking paradigms are supported by various communication standards, communication links, and equipped with different types of network devices. Currently, each communication system is closed due to the proprietary network equipments and the dedicated gateways are required for protocol and format conversions, which significantly limits interoperability. %To fully enable inter-operation, an unified network architecture is needed.

\item Network management: Each network already comprises a large number of devices with dedicated interfaces for configuration and control. The integrated network will become more complex to manage. Moreover, due to the variety of devices with different hardware and software specifications, it is difficult to perform reconfiguration flexibly and dynamically, to either enforce high-level policies, or respond to various events such as network congestion and link failures.

\item Dynamic networking: The mobility of satellites and HAPs with respect to the earth complicates the integrated network operation. For instance, the LEO satellites are spinning around the earth with periods less than 130 minutes\footnote{https://www.n2yo.com/satellites/?c=17, 2017.}, leading to resource availability dynamics. Moreover, vehicle mobility results in time-varying and non-uniform geographical vehicle distribution.  % Additionally, resource usage in each segment is changing dynamically. The resource availability vary since each segment dynamically allocate its resources to satisfy the legacy services (e.g., earth observation in satellite networks).% Moreover, considering the fluctuation in links, bursty vehicular traffic, ....

\item QoS provisioning:
    Integrated networks should accommodate a wide range of vehicular services with different QoS requirements efficiently. However, the resources in the integrated network exhibit high heterogeneity. Moreover, the resource availability for vehicular services varies over time, since each segment dynamically allocates resources to support its legacy services\footnote{Legacy services refer to the existing services in different segments, rather than the vehicular services, e.g., earth observation and navigation in satellite networks.} in priority. %Moreover, various vehicular scenarios should be accommodated in an efficient way, such as highway, urban, rural, and high-speed trains.

\end{itemize}

\section{Software Defined Space-Air-Ground Integrated Vehicular Networks} \label{sec.SDN}
As an emerging network architecture, SDN separates the control plane and data plane, introduces a logically centralized control with a global view of the network, and facilitates network programmability/reconfiguration through open interfaces. With SDN, a dynamic, manageable, cost-effective, and adaptable network can be enabled\footnote{https://www.opennetworking.org/sdn-resources/sdn-definition, 2013}. In this section, based on SDN, we propose a software defined space-air-ground integrated vehicular (SSAGV) network architecture, to address the aforementioned challenges.
\subsection{Network Architecture}

As shown in Fig.~\ref{fig:ssag}, the SSAGV network comprises three main segments: space, air, and ground segments. SDN controllers can be deployed on powerful servers or in cloud computing, which regulate the network behaviors and manage network resources dynamically. Considering different segments have distinct characteristics such as communication standards and diverse network devices with various functions, the control and communication interfaces of SDN controllers for each segment should be dedicated to the corresponding segment. As a fact, software defined paradigms for satellite, aerial, and territorial networks have been proposed separately \cite{bertaux2015software}\cite{iqbal2016software}\cite{pentikousis2013mobileflow}. To orchestrate the operation of each segment, higher-tier SDN controllers are introduced on the top of SDN controllers in each segment to support vehicular services. To facilitate the decision making at different tiers of SDN controllers, different levels of abstracted information regarding the network status are needed, such as vehicle geographical density, location-dependent content popularity, and the number of active vehicular users in a local area.

Vehicular services should not interfere with the legacy services in different segments. To this end, network slicing is performed in each segment to partition the whole network resource into various slices for different services, whereby those slices operate in an isolated manner and do not interfere with each other. To accomplish this goal, hypervisors are integrated into each segment \cite{akyildiz2015softair}. Specifically, the lower-level hypervisors at network components such as RSUs and SBSs schedule the local resources to different slices, where each slice can exclusively use the resources for a certain time period. The upper-level network hypervisor coordinates the lower-level hypervisors to perform network-wide resource allocation. As a result, sets of resources are allocated for the legacy services. For instance, different sets of resources in satellite networks are allocated for earth observation, navigation, and weather monitoring, respectively. Then, the remaining resources from each segment are put into a space-air-ground resource pool for vehicular services. Note that network slicing is dynamically performed to achieve high resource utilization while supporting the time-varying needs of legacy and vehicular services.

\subsection{Layered Structure}
\begin{figure*}[t!]
    \centering
    \includegraphics[width=15cm]{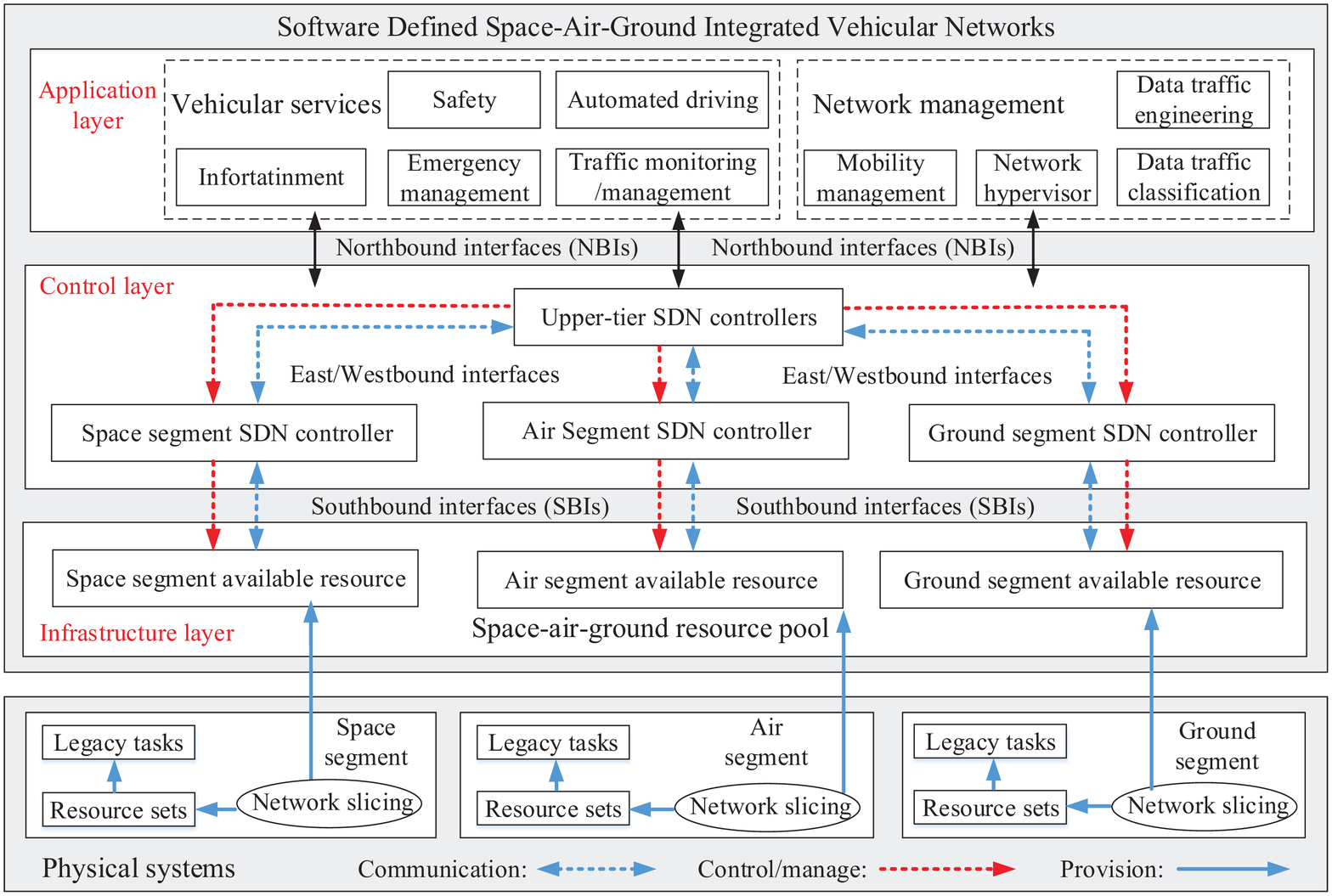}
    \caption{Layered structure of SSAGV networks.}
    \label{fig:layer}
\end{figure*}

As shown in Fig.~\ref{fig:layer}, the proposed SSAGV network is organized in three layers: infrastructure, control, and application layers. In the infrastructure layer, computing, storage, and communication resources from different segments constitute a common resource pool. Since in each segment, network slicing is performed dynamically, the space-air-ground resource pool varies with time. To deal with the dynamics in resource availability, resource virtualization can be adopted to abstract logical resource from the physical resources and vehicular services can be deployed on a collection of virtualized resources, instead of specific physical resources. By dynamically performing virtual and physical resource mapping, disturbance to vehicular services can be alleviated.

In the control layer, for scalability, SDN controllers are organized in a hierarchical manner, targeting at network operation in different domains (i.e., local, regional, national, and global domains). Through southbound interfaces (SBIs), SDN controllers communicate and control the respective underlaying physical resources. Considering the significant heterogeneity in different segments, the implementation of southbound interfaces will be specific to each segment, to facilitate various functions such as beam steering of satellites, movement control of UAV, and resource block allocation in LTE base stations. The implementation of SDN controllers should also address the scope and granularity of control, e.g., an SDN controller can fully or partially control the underlying network components. For partial control, the devices can have local intelligence. To efficiently use heterogenous network resources and conduct different levels of network functions, SDN controllers in different segments are coordinated by upper-tier controllers, through Eastbound and Westbound interfaces.

In the application layer, a variety of vehicular services and network management functions are performed based on the functions provided by the control layer through the northbound interfaces (NBIs). The requests from an application can be translated into rules by NBIs for the SDN controllers, which are further interpreted into instructions to guide the underlaying devices through SBIs. Vehicular services correspond to the services provisioned directly to vehicular users, while network management functions facilitate efficient operation, such as mobility management, network hypervisor, data traffic classification, and data traffic engineering. Specifically, mobility management aims to provide seamless connectivity to moving vehicles by associating them with suitable access points. Data traffic classification associates data traffic flows with different vehicular services and categorize them into different QoS classes, by means of deep packet inspection (DPI) and machine learning. Based on classification, differentiated services can be enabled, e.g., through enforcing corresponding policies. Data traffic engineering can either steer data traffic to different segments and access points (e.g, RSUs and SBSs) for load balancing, or dynamically changes the route to avoid congestion. Network hypervisors can help create multiple virtual networks coexisting over the common physical infrastructure by dynamically scheduling multi-dimensional resources. The virtual networks can be tailored to better support different vehicular services.

With the proposed SSAGV network architecture, the following benefits can be achieved: i) simplified network management and cost-effective network upgrade/evolution; ii) optimized the network operation and resource utilization; and iii) flexible and agile network behavior control on the fly, along with adaption to network dynamics such as topology changes and link failures. The SSAGV network can pave the way towards an open ecosystem with network agility and flexibility, and help achieve efficient interoperability, simplified operation and maintenance, and round-the-clock optimization of network.

\section{Network Operation}\label{sec.operation}

\subsection{Working Relation}

\begin{figure}[t!]
    \centering
    \includegraphics[width=8cm]{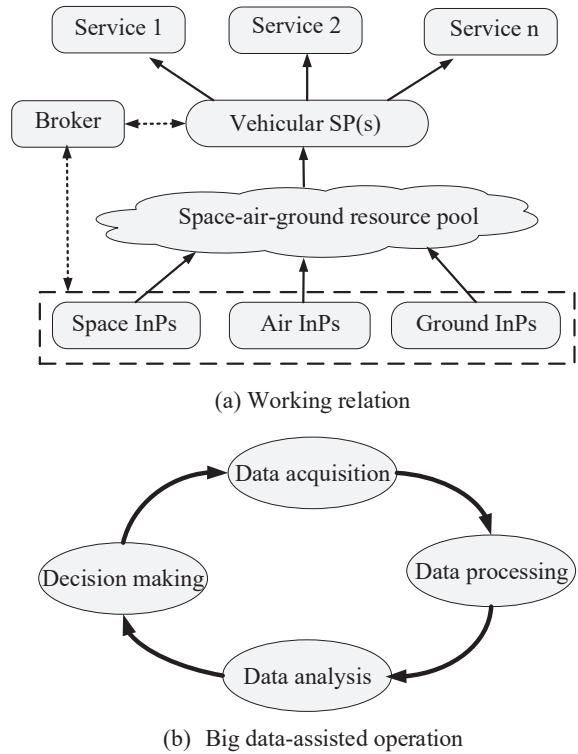}
    \caption{Working relation and big data-assisted operation.}
    \label{fig:business}
\end{figure}

In the SSAGV network, three different parties arise, including infrastructure providers (InPs), vehicular service providers (SPs), and vehicular customers, as shown in Fig.~\ref{fig:business}(a). InPs provide infrastructure resources to vehicular SPs, which then provision diverse services to the vehicular customers. Based on the infrastructure types, InPs mainly encompasses the space segment InPs such as Intelsat, the air segment InPs like Google, and ground segment InPs such as cellular network operators. Vehicular SPs request resources from different InPs to support their vehicular users, while InPs lease their resources considering the payments from vehicular SPs and the requirements of their legacy services. When an InP fails to provision the required resources, coalitions can be formed to satisfy the requirements of vehicular SPs, e.g., different owners of HAPs can negotiate and cooperate to provide the required capacity for a targeted area. To facilitate the interactions between vehicular SPs and InPs, brokers can be introduced, which receive resource requirements from vehicular SPs and negotiate for resources from multiple InPs.
By doing so, resources among different systems can be shared dynamically on demand.
\subsection{Hierarchical Network Operation}
The SSAGV network operation is complex, due to the large network scale, the wide range of services, the heterogenous devices/resources, and high dynamics in network states. Therefore, network operation is performed in a hierarchical manner. Specifically, in the spatial dimension, the network is divided into different domains, which are controlled/managed by the hierarchical SDN controllers correspondingly. Lower-tier SDN controllers manage the underlying network devices in a small area. The upper-tier SDN controllers cover a large area, and can coordinate multiple lower-tier SDN controllers to perform  high-level network operations. Lower-tier SDN controllers perform fine-grained control, while the upper-tier SDN controllers conduct coarse-grained control. For instance, the former can perform power control and user scheduling for the underlying devices such as BSs and RSUs, while the latter can steer different amounts of vehicular data traffic to different segments, or adjust the ON-OFF modes of BSs and RSUs for energy saving, based on the density of vehicular users. Taking content delivery as another example, the upper-tier SDN controllers update the content cached in the servers or BSs in a specific area, according to the time-varying content popularity, while the lower-tier SDN controllers manipulate the content delivery based on instantaneous users' requests.

Accordingly, in the temporal domain, the network operation is performed in a multi-time scale fashion, to deal with the network dynamics, such as time-varying resource availability and burst vehicular service requests. In a large time scale, upper-tier SDN controllers adjust their strategies, according to the high-level network status (e.g., spatial traffic distribution or content popularity). In a small time scale, the lower-tier SDN controllers update their strategies, adapting to instantaneous vehicular user requests. Within different time scales, the corresponding SDN controllers dynamically optimize their control policies based on the respective abstracted network information. Note that the information provided to various tiers of SDN controllers for decision making are different in terms of spatial and temporal scales. For instance, the density of vehicular users can be required for upper-tier SDN controllers, while the detailed location information of vehicular users are needed at lower-tier SDN controllers.

\subsection{Big Data-Assisted Operation}
The decision making of SDN controllers relies on the information collection and analysis. Considering the scale and volume of information, big data techniques can be adopted~\cite{su2016big}. Through big data analytics, insights can be extracted to guide the decisions of SDN controllers. For instance, data traffic exhibit strong correlative and statistical features in temporal and spatial domains. By extracting the data distribution features, network resources can be better provisioned. The big data-assisted operation cycle is shown in Fig.~\ref{fig:business}(b). Specifically, to acquire vehicle related information, in the ground segment, different approaches can be employed to collect road data, including roadside camera, electromagnetic transducer, and acouste sensor; while in the space or air segments, aerial photography and videos on road information can be collected. After the data collection phase, raw data has to be processed before transmission, such as data compression and data fusion. For instance, the satellite with onboard processing (OBP) can extract useful information before transmission. Then, data mining and machine learning techniques can be employed in the data analysis phase in order to obtain the knowledge about the network states. In addition, accurate prediction such as for spatial-temporal traffic distribution, content popularity and user mobility, can facilitate the optimal decision making.  With big data and SDN, intelligent and automated network operation can be achieved.

\begin{table*}[t]\small
\caption{Simulation Parameters}
\centering
\begin{tabular}{ l | l }
\hline\hline
Number of Satellites & 6 in a plane \\ \hline
Low Earth Orbit Altitude		&  1414 km\\ \hline
Period		& 114$\sim$130 minutes\\ \hline
Inclination Angle   & $0^\circ$ \\ \hline
Altitude of HAP		&  20 km\\ \hline
Minimum Elevation Angle   & $10^\circ$ \\ \hline
Radius of Earth		&  6371 km\\ \hline
\hline
\end{tabular}
\label{netset}
\end{table*}

\section{A Case Study}\label{sec.Study}

In this section, a case study is provided on SDN enabled coordination of satellite and HAPs, aiming at delivering contents to vehicular users efficiently.

\subsection{Network Setting}
Suppose that direct optical links from satellite to the ground is poor due to the weather conditions, and HAPs can be utilized as relays. Each HAP is able to connect one satellite through the free-space optical link and connect the vehicular users on the ground using microwave links, while a satellite with multiple beams can communicate with multiple HAPs. Consider a LEO satellite network of Globalstar system, which operates at the height of $1414$ km, and consists of 48 satellites in 8 planes with different inclination angles. HAPs, at the altitude of $20$ km, are uniformly distributed in the air. HAPs stay relatively stable to the earth, while the LEO satellites are spinning around with periods less than 130 minutes. The parameters are summarized in Table \ref{netset}. Note that the minimum elevation angle for communications is set to $10^\circ$.

\begin{figure*}[t!]
\centering
\subfigure[Average signal strength versus the maximum number of the satellite beams.] {\includegraphics[width=0.45\textwidth]{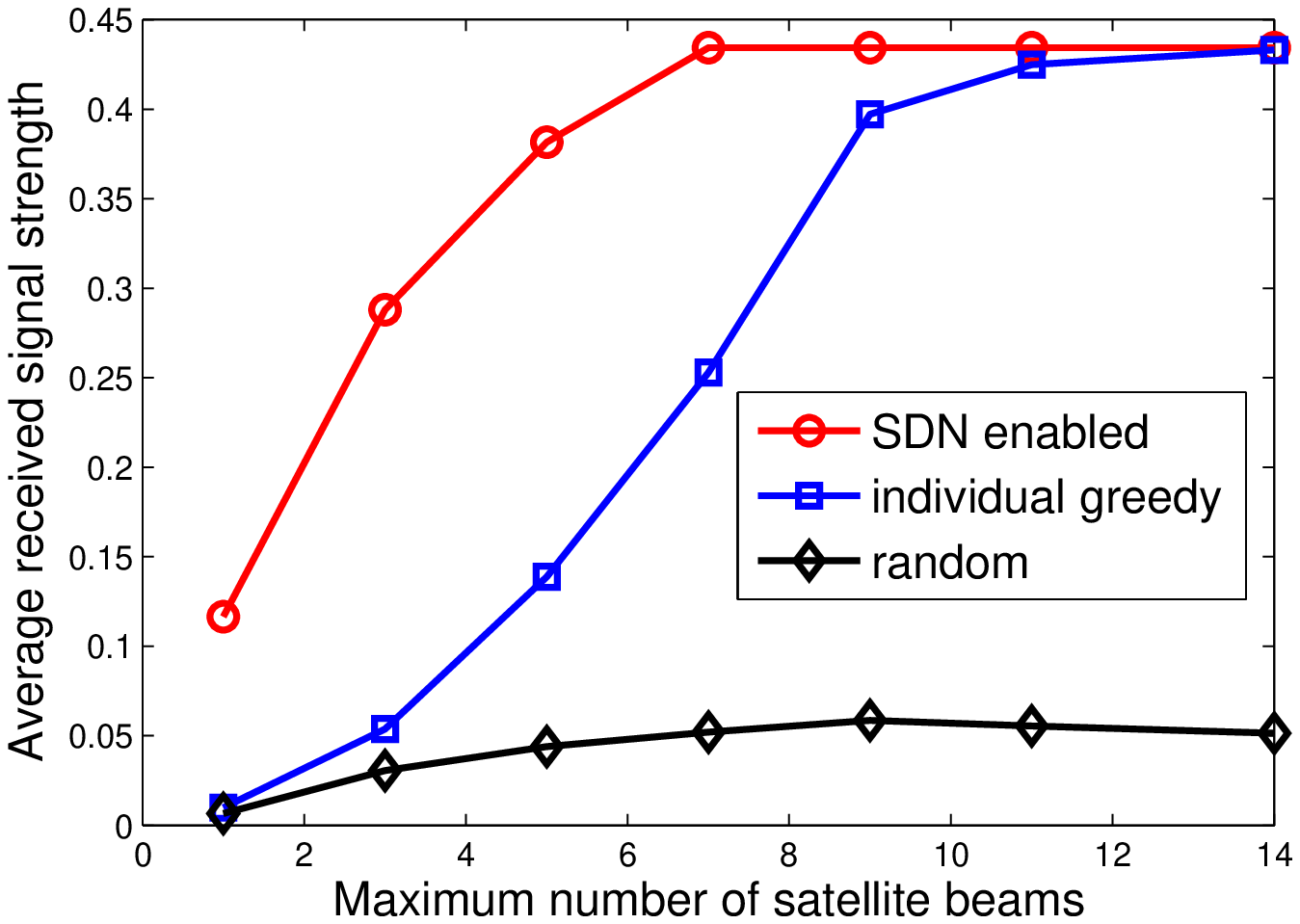} \label{fig:Hmax}}
%\hspace{1in}
\subfigure[Average signal strength versus the number of HAPs.] {\includegraphics[width=0.45\textwidth]{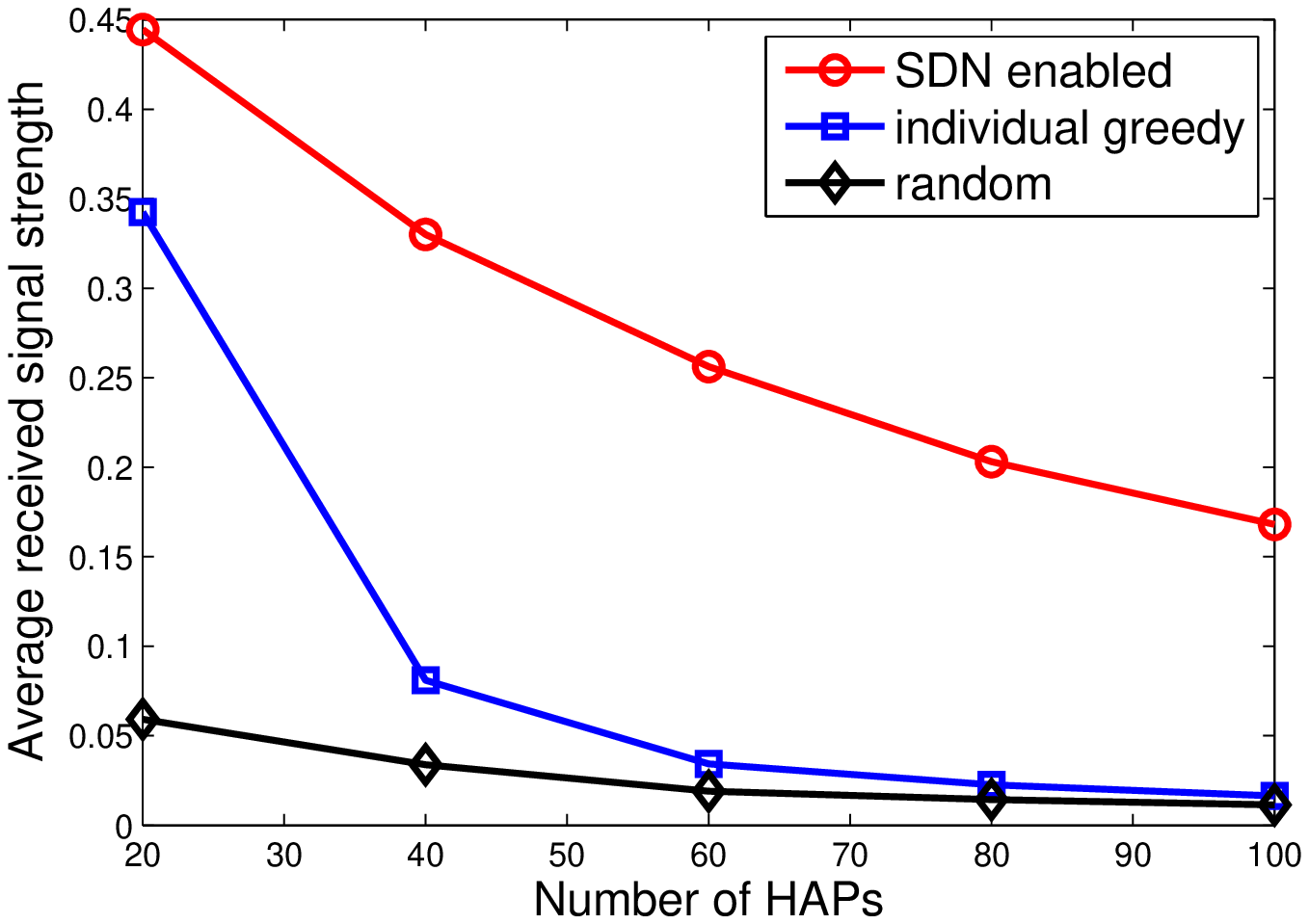} \label{fig:Hap}}
\caption{Performance comparison in terms of average signal strength.}
\end{figure*}

\subsection{Simulation Results}
We aim to maximize the average signal strength at HAPs (equivalent to maximizing the aggregate throughput) by coordinating the connections between HAPs and satellites. The signal strength mainly depends on the elevation angle. Thanks to the control channel, SDN controller has real-time spatial information of both satellites and HAPs, which can be exploited to make around-the-clock optimal connection decisions. Specifically, the connection scheduling problem is a bipartite one-to-many matching between satellites and HAPs; and the SDN enabled coordination scheme determines one-to-many matching between the satellites and HAPs. As comparisons, legacy systems without spatial information can only make myopic decisions by greedily connecting HAPs to satellite beams with a high elevation angle. Results of random link establishment is also presented as a benchmark.

Fig.~\ref{fig:Hmax} shows the average signal strength with respect to the maximum number of the satellite beams, when there are 6 satellites and 50 HAPs. It can be seen that, the SDN-enabled coordination scheme always outperforms individual greedy and random link establishment schemes, and quickly converges to the optimal value. In the individual greedy scheme, each HAP chooses among the available satellites to attain better signal strength. The SDN enabled coordination facilitates a centralized control, and thus helps achieve round-the-clock optimum.

Fig.~\ref{fig:Hap} shows the average signal strength with respect to the number of HAPs, when the maximum link of a satellite is set to 3. It can be seen that the SDN-enabled coordination scheme still outperforms the individual greedy and random link establishment schemes. Moreover, with an increase of number of HAPs, the average signal strength decreases, since more HAPs compete for the limited satellite links.

%When the concurrent links of a satellite is large enough, HAP can communicate to the ones with better connection. In contrast, we show the advantage of our architecture with .

Satellite networks often contain multiple planes, and this study demonstrates the considerable performance gain for one plane. The gain obtained from SDN enabled coordination in a practical network can be even more appealing.

\section{Research Issues}\label{sec.Research}
In this section, some open research issues are discussed for the SSAGV network.

\subsection{Software Defined Platform Implementation}
To exploit the potential benefits of the SSAGV network, a fundamental issue lies in the implementation of the SDN paradigm, including the data plane, the control plane, as well as different open interfaces. Firstly, the scope and the granularity of control for SDN controllers should be carefully devised to balance the performance and complexity. Secondly, since this paradigm spans various segments, the distinct features of each segment should be taken into account, such as the respective control and communication interfaces. Thirdly, the deployment of the hierarchical SDN controllers largely affects the network performance, and it needs further investigation.

\subsection{QoS-aware Resource Allocation}
With different segments integrated in the unified platform, efficient resource allocation becomes a significant challenge to support various vehicular services, due to the highly dynamic network environment and multi-dimensional heterogeneity in resources and services. When performing resource allocation, the characteristics of different segments should be considered, such as the predicted trajectory of satellites and the controllable movement of HAPs. In such a case, a time-evolving resource graph (TERG) can be utilized to describe the evolution of satellite and aerial network resources. The TERG is built on a unified two-dimensional
time-space basis, where vertices correspond to network elements such as satellites and HAPs, and edges denote the availability of different resources. The TERG can be further combined with available resources at the ground to dynamically support vehicular services.

\subsection{Interaction among Different Parties}
A vehicular SP can lease resources from different InPs for vehicular services, where InPs and vehicular SPs have different interests. For instance, the InPs aim to maximize the revenue from the vehicular SP without degrading their own legacy services, whereas the SP attempts to support vehicular services cost-effectively by exploiting different resources from InPs. The interaction between the InPs and the SP greatly affects the service performance, which can be studied through game theory. For example, the interactive procedure of price negotiation and resource dispatch can be modeled and analyzed by non-cooperative games. Furthermore, in the scenarios with multiple InPs and multiple vehicular SPs, the auction theory can be employed to analyze their interactions.

\subsection{Network Virtualization}
To better support multifarious vehicular services/scenarios, network virtualization can be employed to create service-oriented virtual networks. Based on the service requirements, the virtual networks can be customized, including the virtual network topology and end-to-end protocols. Moreover, virtual networks can have stable virtual network topology by dynamically embedding the virtual nodes to the physical nodes, helping mitigate the adverse effects caused by the time-varying physical topology. To facilitate network virtualization, optimal virtual network topology, dynamic network embedding, and customized end-to-end protocols need extensive investigation.

\subsection{Network Security}
Since SDN controllers are responsible for managing resources and controlling network operation, it is imperative to protect the SDN controllers from different cyber attacks. For instance, denial-of-service (DoS) attacks can be launched to paralyze the  operations of SDN controllers, or the SDN controllers can be compromised through inside attacks. In addition, variety of vehicular services require different security goals in terms of confidentiality, integrity, and authentication. Those requirements translate into various quality of protection (QoP) parameters such as cryptographic variables and encryption key lengths. To better protect the network and vehicular services, the security aspect of SSAGV networks should be investigated.

\section{Conclusion}\label{sec.Conclusion}
In this article, we have proposed an SSAGV network architecture to exploit the advantages of space, air, and ground segments, to support diverse vehicular services in various scenarios efficiently and cost-effectively. The proposed open network architecture can achieve network agility and flexibility, simplify network management and maintenance, and adapt to changing user demands and
network states. The new network operation model can pave the road for resource sharing and collaboration among different segments. To accelerate the pace of SSAGV network development, extensive research efforts are required in the outlined research
directions.

\vspace*{-1.5\baselineskip}
\begin{IEEEbiography} {Ning Zhang}
(S'12,M'16) earned the Ph.D degree from University of Waterloo in 2015. He received his B.Sc. degree from Beijing Jiaotong University and the M.Sc. degree from Beijing University of Posts and Telecommunications, Beijing, China, in 2007 and 2010, respectively. From May 2015 to April 2016, he was a postdoc research fellow at BBCR lab in University of Waterloo. He is now a postdoc research fellow at University of Toronto. He is now an associate editor of International Journal of Vehicle Information and Communication Systems and a guest editor of Mobile Information System. He is the recipient of the Best Paper Award at IEEE Globecom 2014 and IEEE WCSP 2015. His current research interests include next generation wireless networks, software defined networking, network virtualization, and physical layer security.
\end{IEEEbiography}

\begin{IEEEbiography}
{Shan Zhang} received her Ph.D. degree in Department of Electronic Engineering from Tsinghua University and B.S. degree in Department of Information from Beijing Institute Technology, Beijing, China, in 2016 and 2011, respectively. She is currently a post-doctoral fellow in Department of Electrical and Computer Engineering, University of Waterloo, Ontario, Canada. Her research interests include resource and traffic management for green communication, intelligent vehicular networking, and software defined networking. Dr. Zhang received the Best Paper Award at the Asia-Pacific Conference on Communication in 2013.
\end{IEEEbiography}

\begin{IEEEbiography}
{Peng Yang} received his B.Sc. degree from the Department of Electronics and Information Engineering, Huazhong University of Science and Technology (HUST), Wuhan, China, in 2013. Currently, he is pursuing his Ph.D. degree in the School of Electronic Information and Communications, HUST. From Sept. 2015, he is also a visiting Ph.D. student in the Department of Electrical and Computer Engineering, University of Waterloo, Canada. His current research interests include next generation wireless networking, software defined networking and fog computing.
\end{IEEEbiography}

\begin{IEEEbiography}
{Omar Alhussein} (S'14) received the B.Sc. degree in communications engineering from Khalifa University, Abu Dhabi, United Arab Emirates, in 2013 and the M.A.Sc. degree in engineering science from Simon Fraser University, Burnaby, BC, Canada, in 2015. He is currently working towards the Ph.D. degree in the Broadband Communications Research Laboratory, University of Waterloo, Waterloo, ON, Canada. From January 2014 to May 2014, he was a Research Assistant with the Etisalat BT Innovation Centre (EBTIC), Khalifa University. From May 2014 to September 2015, he was with the Multimedia Communications Laboratory, Simon Fraser University. His research interests are in areas spanning software defined networking, network virtualization, wireless communications, and machine learning. Mr. Alhussein currently serves as a reviewer for IEEE Communications Letters, IEEE Transactions ON Vehicular Technology, and other journals and conferences.
\end{IEEEbiography}

\begin{IEEEbiography} {Weihua Zhuang} (M'93-SM'01-F'08) has been with the Department of Electrical and Computer Engineering,
University of Waterloo, Canada, since 1993, where she is a Professor and a Tier I Canada Research Chair in Wireless Communication Networks. Her current research focuses on resource allocation and QoS provisioning in wireless networks, and on smart grid. She is a co-recipient of several best paper awards from IEEE conferences. Dr. Zhuang was the Editor-in-Chief of IEEE Transactions on Vehicular Technology (2007-2013), and the TPC Co-Chair of IEEE VTC Fall 2016. She is a Fellow of the IEEE, a Fellow of the Canadian Academy of Engineering, a Fellow of the Engineering Institute of Canada, and an elected member in the Board of Governors and VP Publications of the IEEE Vehicular Technology Society.
\end{IEEEbiography}

\begin{IEEEbiography}{Xuemin (Sherman) Shen} (M'97, SM'02, F'09)
received the B.Sc. (1982) degree from
Dalian Maritime University (China) and the M.Sc.
(1987) and Ph.D. degrees (1990) from Rutgers
University, New Jersey (USA), all in electrical
engineering. He is a Professor and University
Research Chair, Department of Electrical
and Computer Engineering, University of Waterloo,
Canada. He was the Associate Chair for Graduate
Studies from 2004 to 2008. Dr. Shen¡¯s research
focuses on resource management in interconnected
wireless/wired networks, wireless network security, social networks, smart
grid, and vehicular ad hoc and sensor networks. Dr. Shen served as the
Technical Program Committee Chair/Co-Chair for IEEE Infocom¡¯14, IEEE
VTC¡¯10 Fall, the Symposia Chair for IEEE ICC¡¯10, the Tutorial Chair for
IEEE VTC¡¯11 Spring and IEEE ICC¡¯08, the Technical Program Committee
Chair for IEEE Globecom¡¯07, the General Co-Chair for Chinacom¡¯07 and
QShine¡¯06, the Chair for IEEE Communications Society Technical
Committee on Wireless Communications, and P2P Communications and
Networking. He also serves/served as the Editor-in-Chief for IEEE Network,
Peer-to-Peer Networking and Application, and IET Communications; a
Founding Area Editor for IEEE Transactions on Wireless Communications;
an Associate Editor for IEEE Transactions on Vehicular Technology,
Computer Networks, and ACM/Wireless Networks, etc.; and the Guest
Editor for IEEE JSAC, IEEE Wireless Communications, IEEE
Communications Magazine, and ACM Mobile Networks and Applications,
etc. Dr. Shen received the Excellent Graduate Supervision Award in 2006,
and the Outstanding Performance Award in 2004, 2007 and 2010 from the
University of Waterloo, the Premier¡¯s Research Excellence Award (PREA)
in 2003 from the Province of Ontario, Canada, and the Distinguished
Performance Award in 2002 and 2007 from the Faculty of Engineering,
University of Waterloo. Dr. Shen is a registered Professional Engineer of
Ontario, Canada, an IEEE Fellow, an Engineering Institute of Canada
Fellow, a Canadian Academy of Engineering Fellow, a Royal Society of
Canada Fellow, and a Distinguished Lecturer of IEEE Vehicular Technology
Society and Communications Society.
\end{IEEEbiography}

\end{document}